\documentclass[11pt,fleqn]{iopart}
\usepackage[utf8]{inputenc}
\usepackage[T1]{fontenc}
\usepackage{fouriernc}
\expandafter\let\csname equation*\endcsname=\relax
\expandafter\let\csname endequation*\endcsname=\relax
\usepackage{amsmath,amssymb,amsfonts,amsthm}
\usepackage{hyperref,xcolor}
\usepackage{etoolbox}
\usepackage{units}

\numberwithin{equation}{section}

\newenvironment{centeqn}{
 \makeatletter
 \setbool{@fleqn}{false}
 \makeatother
 \begin{equation}}
 {\end{equation}
}

\def\osp{{\mathfrak{osp}}}

\def\sl{{\mathfrak{sl}}}
\def\os{{\mathfrak{o}}}

\def\bo{{\mathcal{O}}}

\def\dir{{\underline{D}}}
\def\xir{{\underline{x}}}
\def\euler{{\mathbb{E}}}
\def\inv{{\mathcal{S}}}
\def\cliff{{C\!\!\ell}}

\newcommand{\obar}[1]{\check{#1}}
\newcommand{\olin}[1]{\overline{#1}}

\begin{document}

\title[The dual pair $Pin(2n)\times\mathfrak{osp}(1|2)$, the Dirac equation and the Bannai-Ito algebra]{The dual pair $Pin(2n)\times\mathfrak{osp}(1|2)$, the Dirac equation and the Bannai-Ito algebra}

\author{Julien Gaboriaud}
\ead{gaboriaud@CRM.UMontreal.CA}
\address{Centre de Recherches Math\'ematiques, Universit\'e de Montr\'eal, Montr\'eal (QC), Canada}

\author{Luc Vinet}
\ead{vinet@CRM.UMontreal.CA}
\address{Centre de Recherches Math\'ematiques, Universit\'e de Montr\'eal, Montr\'eal (QC), Canada}

\author{St\'ephane Vinet}
\ead{stephanevinet@uchicago.edu}
\address{The College, The University of Chicago, 5801 S. Ellis Ave, Chicago, IL 60637, USA}

\author{Alexei Zhedanov}
\ead{zhedanov@yahoo.com}
\address{Department of Mathematics, Renmin University of China, Beijing 100872, China}

\vspace{10pt}

\begin{abstract}   
The Bannai-Ito algebra can be defined as the centralizer of the coproduct embedding of $\mathfrak{osp}(1|2)$ in $\mathfrak{osp}(1|2)^{\otimes n}$. It will be shown that it is also the commutant of a maximal Abelian subalgebra of $\mathfrak{o}(2n)$ in a spinorial representation and an embedding of the Racah algebra in this commutant will emerge. The connection between the two pictures for the Bannai-Ito algebra will be traced to the Howe duality which is embodied in the $Pin(2n)\times\mathfrak{osp}(1|2)$ symmetry of the massless Dirac equation in $\mathbb{R}^{2n}$. Dimensional reduction to $\mathbb{R}^{n}$ will provide an alternative to the Dirac-Dunkl equation as a model with Bannai-Ito symmetry. 
\end{abstract}

\section{Introduction}
The Bannai-Ito algebra $B(n)$ can be presented in terms of generators and relations \cite{DeBie2016a,DeBie2017}. Let $[n]=\{1,2,\dots,n\}$ denote the set of the $n$ first integers and \mbox{$S=\{s_1,\dots,s_k\}$} be an ordered $k$-subset of $[n]$. The generators $\Gamma_S$ of $B(n)$ are labelled by all subsets for $k=0,1,\dots,n$. For any two subsets $A$ and $B$ of $[n]$, the relations between the generators $\Gamma_A$ and $\Gamma_B$ that define $B(n)$ are:
\begin{align}\label{eq_BIn}
 \{\Gamma_{A},\Gamma_{B}\}=\Gamma_{(A\cup B)\setminus(A\cap B)}+2\Gamma_{A\cap B}\Gamma_{A\cup B}+2\Gamma_{A\setminus(A\cap B)}\Gamma_{B\setminus(A\cap B)},
\end{align}
where $\{X,Y\}=XY+YX$. By convention $\Gamma_{\emptyset}=-\nicefrac{1\,\,}{\,2}$, and the generators associated to a set are simply labelled by the indices $\Gamma_{\{i_1,\dots,i_k\}}\equiv\Gamma_{i_1\,\dots\,i_k}$. Moreover, $\Gamma_{i}$, $i\in[n]$ and $\Gamma_{[n]}$ are central.

For the rank one case which occurs when $n=3$, the relations for $B(3)$ are seen to be \cite{DeBie2016a,DeBie2015}
\begin{align}\label{eq_BI3}
 \{\Gamma_{ij},\Gamma_{jk}\}=\Gamma_{ik}+2\Gamma_{j}\Gamma_{ijk}+2\Gamma_{i}\Gamma_{k},
\end{align}
where $i,j,k\in[3]$ are all distinct.

We shall present in this paper a defining context for $B(n)$ (and $B(3)$) as a commutant and relate this result to a Howe duality framework.

The Bannai-Ito algebra $B(3)$ was initially introduced in \cite{Tsujimoto2012} to encode the bispectral properties of the Bannai-Ito polynomials which were discovered by the researchers whose name they bear in a classification problem in algebraic combinatorics \cite{Bannai1984}. $B(3)$ was later seen \cite{Genest2016} to be isomorphic to a degenerate double affine Hecke algebra (DAHA) of type $(C_1^{\vee},C_1)$. A key connection between the Lie superalgebra $\osp(1|2)$ and the Bannai-Ito algebra was also made \cite{DeBie2015}. Indeed, the Bannai-Ito polynomials were identified as forming the Racah coefficients for $\osp(1|2)$ \cite{Genest2014b}. It was then shown quite naturally that $B(3)$ is realized by the intermediate Casimir operators arising in the threefold tensor product of $\osp(1|2)$ with itself. Extending this construction to the $n$-fold tensor product $\osp(1|2)^{\otimes n}$ led to the definition of $B(n)$ \cite{DeBie2016a}. (More details will be given on this later.) A number of applications for $B(n)$ were subsequently found. For instance, the conserved quantities of a superintegrable model on the $(n-1)$-sphere \cite{Genest2015a,Genest2014c,DeBie2016b,DeBie2017b} were seen to satisfy the commutation relations \eqref{eq_BIn}. Of particular relevance to the present study is the fact that $B(n)$ is also the symmetry algebra of the (massless) Dirac-Dunkl \cite{DeBie2016a,DeBie2016} equation in $\mathbb{R}^{n}$.

The Bannai-Ito algebra $B(n)$ has much in common with the Racah algebra $R(n)$ \cite{DeBie2017a}. Like $B(n)$, $R(n)$ can be defined in a tensorial fashion as the algebra formed by the intermediate Casimir operators associated to the $n$-fold tensor product of the Lie algebra $\sl(2)$ with itself. $R(n)$ is also the symmetry algebra of a certain generic superintegrable model on $S^{n-1}$ (without reflections). The relation between the rank one version $R(3)$ and the Racah polynomials \cite{Koekoek2010} is quite similar to the one between the Bannai-Ito algebra $B(3)$ and the Bannai-Ito polynomials. $R(3)$ has three generators $K_1$, $K_2$, $K_3$ which are subjected to the relations
\begin{align}\label{eq_racah3}
\begin{aligned}{}
 [K_1,K_2]=K_3 \qquad\quad\  [K_2,K_3]&={K_2}^2+\{K_1,K_2\}+dK_2+e_1 \quad\ \\
 [K_3,K_1]&={K_1}^2+\{K_1,K_2\}+dK_1+e_2
\end{aligned}
\end{align}
with $d$, $e_1$, $e_2$ central. The defining relations for $R(n)$ \cite{DeBie2017a} are given in Section \ref{sec_embedding} where an embedding of $R(n)$ into $B(n)$ will be identified.

We have recently observed that $R(3)$ is the commutant in oscillator representations of the universal enveloping algebra $\mathcal{U}(\os(6))$ of the subalgebra \mbox{$\os(2)\oplus\os(2)\oplus\os(2)$} of the Lie algebra $\os(6)$ of rotations in six dimensions \cite{Gaboriaud2018}. This was extended to $R(n)$ in \cite{Gaboriaud2018a}. We shall provide here an analogous description of $B(n)$ and shall focus first on $B(3)$.

After recalling in Section \ref{sec_usualosp} how $B(3)$ can be viewed as the centralizer of the coproduct embedding of $\osp(1|2)$ into $\osp(1|2)^{\otimes3}$, we shall show in Section \ref{sec_B3comm} that $B(3)$ is the commutant of \mbox{$\os(2)\oplus\os(2)\oplus\os(2)$} in the enveloping algebra of the spinorial representation of $\os(6)$ associated to the Clifford algebra in $\mathbb{R}^{6}$. By considering the (massless) Dirac equation, a Howe duality setting will be brought up to explain the connection between the results of the two previous sections. Dimensional reduction will be performed in Section \ref{sec_dimred} to complete the picture and will result in a new class of model with Bannai-Ito symmetry. In Section \ref{sec_higher} the generalization of these results to the Bannai-Ito algebra $B(n)$ will be obtained for $n>3$. As mentioned above, an embedding of the higher rank $R(n)$ Racah algebra in the $B(n)$ Bannai-Ito algebra will be explicitly given in Section \ref{sec_embedding}, thus linking the construction presented here to the one in \cite{Gaboriaud2018a}. Brief concluding remarks will follow. Finally, in \ref{sec_appendix} we indicate that our results offer as a byproduct a derivation through dimensional reduction of the superconformal quantum Hamiltonian of Fubini and Rabinovici which is known to be invariant under $\sl(2|1)$ \cite{Fubini1984}.

\section{The superalgebra $\osp(1|2)$ and the Bannai-Ito algebra as a centralizer}\label{sec_usualosp}
The superalgebra $\osp(1|2)$ can be presented as follows. Let $J_0$, $J_{\pm}$ be respectively the even and odd generators of the algebra, obeying the relations
\begin{subequations}\label{eq_osp12}
\begin{gather}
 [J_0,J_\pm]=\pm J_\pm, \qquad\quad \{J_+,J_-\}=2J_0.
\end{gather}
The $\mathbb{Z}_2$-grading of the superalgebra can be encoded through a grading involution $\inv$, which commutes with even elements and anticommutes with odd elements:
\begin{gather}
 [\inv,J_0]=0,\qquad\quad \{\inv,J_\pm\}=0.
\end{gather}
\end{subequations}
The sCasimir operator of $\osp(1|2)$ given by
\begin{align}
 S=\frac{1}{2}\left([J_-,J_+]-1\right)
\end{align}
commutes with all the odd elements and anticommutes with all the even ones, that is \mbox{$[S,J_0]=0$}, \mbox{$\{S,J_\pm\}=0$}. It is then straightforward to define a Casimir of $\osp(1|2)$ by combining the sCasimir with the involution:
\begin{align}\label{eq_Cosp}
 \Gamma=S\,\inv=\frac{1}{2}\left([J_-,J_+]-1\right)\inv.
\end{align}
This Casimir $\Gamma$ commutes with all the elements of $\osp(1|2)$.\\
There is a coassociative algebra morphism, the coproduct $\Delta$, which acts as follows:
\begin{align}\label{eq_coproductosp}
 &~\Delta:\osp(1|2)\to\osp(1|2)\otimes\osp(1|2)\nonumber\\
 &\begin{aligned}
  \Delta(J_0)&=J_0\otimes1+1\otimes J_0&&=J_0^{(1)}+J_0^{(2)},\\
  \Delta(J_\pm)&=J_\pm\otimes\inv+1\otimes J_\pm&&=J_\pm^{(1)}\inv^{(2)}+J_\pm^{(2)},\\
  \Delta(\inv)&=\inv\otimes\inv&&=\inv^{(1)}\inv^{(2)},
 \end{aligned}
\end{align}
where the superindex denotes on which factor of the tensor product the generator is acting. 

Now consider the product of three copies of $\osp(1|2)$. The generators corresponding to embeddings in two factors are
\begin{align}
\begin{aligned}
 &\hspace{4em} J_0^{(ij)}=J_0^{(i)}+J_0^{(j)},\qquad \inv^{(ij)}=\inv^{(i)}\inv^{(j)},\qquad i,j=1,2,3,\\
 &J_{\pm}^{(12)}=J_\pm^{(1)}\inv^{(2)}+J_{\pm}^{(2)},\quad J_{\pm}^{(23)}=J_\pm^{(2)}\inv^{(3)}+J_{\pm}^{(3)},\quad J_{\pm}^{(13)}=J_\pm^{(1)}\inv^{(2)}\inv^{(3)}+J_{\pm}^{(3)}.                                                                                                                                                                                                                                                                                                                                                                                                                                                                                                                                                    
\end{aligned}
\end{align}
Note the presence of $\inv^{(2)}$ in $J_{\pm}^{(13)}$. Applying the coproduct twice yields
\begin{align}\label{eq_coproduct2osp}
 \begin{aligned}
  \Delta^{(2)}(J_0)&=J_0^{(1)}+J_0^{(2)}+J_0^{(3)}=J_0^{(123)},\\
  \Delta^{(2)}(J_\pm)&=J_\pm^{(1)}\inv^{(2)}\inv^{(3)}+J_\pm^{(2)}\inv^{(3)}+J_\pm^{(3)}=J_\pm^{(123)},\\
  \Delta^{(2)}(\inv)&=\inv^{(1)}\inv^{(2)}\inv^{(3)}=\inv^{(123)},\qquad\qquad\\\text{with  ~}~ &\Delta^{(2)}=(\Delta\otimes1)\circ\Delta.
 \end{aligned}
\end{align}
The intermediate Casimirs $\Gamma_A$ associated to embeddings of $\osp(1|2)$ in $\osp(1|2)^{\otimes 3}$ labelled by $A\subset[3]$ can then be obtained from the sets above:
\begin{align}
\begin{aligned}
 \Gamma_{i}&=\frac{1}{2}\left([J_-^{(i)},J_+^{(i)}]-1\right)\inv^{(i)},\\
 \Gamma_{ij}&=\frac{1}{2}\left([J_-^{(ij)},J_+^{(ij)}]-1\right)\inv^{(ij)},
\end{aligned}\qquad
 \Gamma_{123}&=\frac{1}{2}\left([J_-^{(123)},J_+^{(123)}]-1\right)\inv^{(123)}.
\end{align}
Even though the intermediate Casimirs $\Gamma_{A}$ commute with the action of $\osp(1|2)$, they do not all commute with each other. 
A direct computation shows that they precisely obey the commutation relations \eqref{eq_BI3} of $B(3)$ thereby proving that this algebra is the centralizer of $\Delta^{(2)}(\osp(1|2))$ in $\mathcal{U}(\osp(1|2)^{\otimes3})$.

\section{The rank $1$ Bannai-Ito algebra as a commutant}\label{sec_B3comm}
We will now show how the Bannai-Ito algebra arises as the commutant of the subalgebra \mbox{$\os(2)\oplus\os(2)\oplus\os(2)$} of $\os(6)$ in spinorial representations of $\mathcal{U}(\os(6))$.

Let $\cliff_6$ be the Clifford algebra generated by the elements\, $\gamma_1,\dots,\gamma_6$\, verifying the relations
\begin{align}\label{eq_clifford}
 \{\gamma_\mu,\gamma_\nu\}=-2\delta_{\mu\nu}.
\end{align}
Denote by $\ell_{\mu\nu}$,~ $\mu,\nu=1,\dots,6$\, the generators of $\os(6)$ which obey
\begin{align}\label{eq_Lijrel}
 [\ell_{\mu\nu},\ell_{\rho\sigma}]=\delta_{\nu\rho}\ell_{\mu\sigma}-\delta_{\nu\sigma}\ell_{\mu\rho}-\delta_{\mu\rho}\ell_{\nu\sigma}+\delta_{\mu\sigma}\ell_{\nu\rho}.
\end{align}
We shall consider the following representation of the algebra of $Pin(6)$ where
\begin{align}
 J_{\mu\nu}&=-iL_{\mu\nu}+\frac{1}{2}\Sigma_{\mu\nu},\label{eq_Jij}
\end{align}
with
\begin{align}
 L_{\mu\nu}&=x_\mu\frac{\partial}{\partial x_\nu}-x_\nu\frac{\partial}{\partial x_\mu},\label{eq_Lij}
\end{align}
and $\Sigma_{\mu\nu}=i\gamma_\mu \gamma_\nu$ the spin operators.

We are interested in the commutant of the \mbox{$\os(2)\oplus\os(2)\oplus\os(2)$} subalgebra of $\os(6)$ represented by the set of elements $\{J_{12},J_{34},J_{56}\}$. First note that:
\begin{align}
 [J_{\mu\nu},L_{\mu\rho}\Sigma_{\mu\rho}+L_{\nu\rho}\Sigma_{\nu\rho}]=0.
\end{align}
Remark also that $[J_{\mu\nu},\Sigma_{\mu\nu}]=0$ and finally that $[J_{\mu\nu},\Sigma_{\rho\sigma}]=0$ for $\mu,\nu,\rho,\sigma$ all different.

It is observed that the commutant of the span of $\{J_{12},J_{34},J_{56}\}$ is generated by the operators (see the remark below):
\begin{align}
\begin{aligned}
 \hspace{-1em}M_{1}&=\left(L_{12}\gamma_1\gamma_2+L_{13}\gamma_1\gamma_3+L_{14}\gamma_1\gamma_4+L_{23}\gamma_2\gamma_3+L_{24}\gamma_2\gamma_4+L_{34}\gamma_3\gamma_4\right)\Sigma_{12}\Sigma_{34}\\
 \hspace{-1em}M_{2}&=\left(L_{34}\gamma_3\gamma_4+L_{35}\gamma_3\gamma_5+L_{36}\gamma_3\gamma_6+L_{45}\gamma_4\gamma_5+L_{46}\gamma_4\gamma_6+L_{56}\gamma_5\gamma_6\right)\Sigma_{34}\Sigma_{56}\\
 \hspace{-1em}M_{3}&=\left(L_{12}\gamma_1\gamma_2+L_{15}\gamma_1\gamma_5+L_{16}\gamma_1\gamma_6+L_{25}\gamma_2\gamma_5+L_{26}\gamma_2\gamma_6+L_{56}\gamma_5\gamma_6\right)\Sigma_{12}\Sigma_{56}.
\end{aligned}
\end{align}
We can now see that these generate the (rank $1$) Bannai-Ito algebra. It is convenient to first introduce the shortened notation $\olin{j}\equiv\{2j-1,2j\}$,~ $j\in\mathbb{N}$. Take now the elements
\begin{align}\label{eq_Gamma1234}
\begin{aligned}
 \Gamma_{\olin{1}\,\olin{2}}&=M_1+\frac{3}{2}\Sigma_{\olin{1}}\Sigma_{\olin{2}},\\
 \Gamma_{\olin{2}\,\olin{3}}&=M_2+\frac{3}{2}\Sigma_{\olin{2}}\Sigma_{\olin{3}},\qquad\qquad \Gamma_{\olin{j}}=\left(L_{2j-1,2j}\,\gamma_{2j-1}\gamma_{2j}+\frac{1}{2}\right)\Sigma_{2j-1,2j}=J_{2j-1,2j}.\\
 \Gamma_{\olin{1}\,\olin{3}}&=M_3+\frac{3}{2}\Sigma_{\olin{1}}\Sigma_{\olin{3}},
\end{aligned}
\end{align}
A straightforward calculation in the realization \eqref{eq_Lij} shows that one has the defining relations of the rank $1$ Bannai-Ito algebra $B(3)$
\begin{align}\label{eq_commBI}
\begin{aligned}
 \{\Gamma_{\olin{1}\,\olin{2}},\Gamma_{\olin{2}\,\olin{3}}\}&=\Gamma_{\olin{1}\,\olin{3}}+2\Gamma_{\olin{2}}\Gamma_{\olin{1}\,\olin{2}\,\olin{3}}+2\Gamma_{\olin{3}}\Gamma_{\olin{1}},\\
 \{\Gamma_{\olin{2}\,\olin{3}},\Gamma_{\olin{1}\,\olin{3}}\}&=\Gamma_{\olin{1}\,\olin{2}}+2\Gamma_{\olin{3}}\Gamma_{\olin{1}\,\olin{2}\,\olin{3}}+2\Gamma_{\olin{1}}\Gamma_{\olin{2}},\\
 \{\Gamma_{\olin{1}\,\olin{3}},\Gamma_{\olin{1}\,\olin{2}}\}&=\Gamma_{\olin{2}\,\olin{3}}+2\Gamma_{\olin{1}}\Gamma_{\olin{1}\,\olin{2}\,\olin{3}}+2\Gamma_{\olin{2}}\Gamma_{\olin{3}},
\end{aligned}
\end{align}
where $\Gamma_{\olin{1}\,\olin{2}\,\olin{3}}$ denotes the (Casimir) element
\begin{align}\label{eq_Gamma6}
\begin{aligned}
 \Gamma_{\olin{1}\,\olin{2}\,\olin{3}}&=\left(\sum_{1\leq \mu<\nu\leq6}-iL_{\mu\nu}\Sigma_{\mu\nu}+\frac{5}{2}\right)\Sigma_{\olin{1}}\Sigma_{\olin{2}}\Sigma_{\olin{3}}
\end{aligned}
\end{align}
and $\Gamma_{\olin{1}\,\olin{2}\,\olin{3}}$, $\Gamma_{\olin{1}}$, $\Gamma_{\olin{2}}$, $\Gamma_{\olin{3}}$ are all central.\\[1em]
\textit{Proof:} Let us explain how the expression for $\{\Gamma_{\olin{1}\,\olin{2}},\Gamma_{\olin{2}\,\olin{3}}\}$ is derived. (The other anticommutators are obtained in a similar way.) Making use of the Clifford algebra properties \eqref{eq_clifford} as well as the $\os(6)$ commutation relations \eqref{eq_Lijrel}, one obtains
\begin{align}\label{eq_details}
\begin{aligned}
 \hspace{-1em}\{\Gamma_{\olin{1}\,\olin{2}},\Gamma_{\olin{2}\,\olin{3}}\}&=3(L_{12}\gamma_1\gamma_2+2L_{34}\gamma_3\gamma_4+L_{56}\gamma_5\gamma_6)+2L_{12}\gamma_1\gamma_2L_{56}\gamma_5\gamma_6\\
 &+2(L_{15}\gamma_1\gamma_5+L_{16}\gamma_1\gamma_6+L_{25}\gamma_2\gamma_5+L_{26}\gamma_2\gamma_6)+\tfrac{9}{2}\\
 &+L_{13}\gamma_1\gamma_3+L_{14}\gamma_1\gamma_4+L_{23}\gamma_2\gamma_3+L_{24}\gamma_2\gamma_4\\
 &+L_{35}\gamma_3\gamma_5+L_{36}\gamma_3\gamma_6+L_{45}\gamma_4\gamma_5+L_{46}\gamma_4\gamma_6\\
 &+2L_{34}\gamma_3\gamma_4\Big(L_{12}\gamma_1\gamma_2+L_{13}\gamma_1\gamma_3+L_{14}\gamma_1\gamma_4+L_{23}\gamma_2\gamma_3+L_{24}\gamma_2\gamma_4\\
 &\qquad+L_{34}\gamma_3\gamma_4+L_{35}\gamma_3\gamma_5+L_{36}\gamma_3\gamma_6+L_{45}\gamma_4\gamma_5+L_{46}\gamma_4\gamma_6+L_{56}\gamma_5\gamma_6\Big)\\
 &-2(L_{13}L_{45}+L_{14}L_{53})\gamma_3\gamma_4\gamma_1\gamma_5-2(L_{13}L_{46}+L_{14}L_{36})\gamma_3\gamma_4\gamma_1\gamma_6\\
 &-2(L_{23}L_{45}+L_{24}L_{53})\gamma_3\gamma_4\gamma_2\gamma_5-2(L_{23}L_{46}+L_{24}L_{36})\gamma_3\gamma_4\gamma_2\gamma_6.
\end{aligned}
\end{align}
The additional identity \cite{Feigin2015}
\begin{align}
 L_{ab}L_{cd}+L_{ac}L_{db}+L_{ad}L_{bc}=0
\end{align}
satisfied in our realization is then the required tool in order to rewrite \eqref{eq_details} as the r.h.s of \eqref{eq_commBI} using the definitions \eqref{eq_Gamma1234} and \eqref{eq_Gamma6}.\\[1em]
\textit{Remark:} It is fairly obvious that the elements $\{G^{i},K^{ij}\}_{1\leq i<j\leq 3}$
\begin{align}\label{eq_GiKij}
\begin{aligned}
 G^{i}&=L_{2i-1,2i}^{2},\\
 K^{ij}&=L_{2i-1,2i}^2+L_{2i-1,2j-1}^2+L_{2i-1,2j}^2+L_{2i,2j-1}^2+L_{2i,2j}^2+L_{2j-1,2j}^2,
\end{aligned}
\end{align}
also belong to the commutant of $\{J_{12},J_{34},J_{56}\}$. It can be seen that they are algebraically dependent on the Bannai-Ito generators given above. It is interesting to observe however that these $G^{i}$ and $K^{ij}$ realize the Racah algebra $R(3)$ as shown in \cite{Gaboriaud2018}. This highlights the fact that the Racah algebra can be embedded in the Bannai-Ito algebra. The explicit embedding and a generalization to the higher rank algebras will be given in Section \ref{sec_embedding}. Operators $\widetilde{G}^{i}$ and $\widetilde{K}^{ij}$ obtained by replacing $L_{ij}$ by $\Sigma_{ij}$ also belong obviously to the commutant of $\{J_{12},J_{34},J_{56}\}$ but lead to trivial operators.

\section{The Dirac model and Howe duality}
We now wish to shed light on the result of the previous two sections by casting in a Howe duality context the observation that the Bannai-Ito algebra arises as the commutant of both $\osp(1|2)$ in $\mathcal{U}(\osp(1|2)^{\otimes 3})$ and $\os(2)^{\oplus 3}$ in the considered spinorial representations of $\mathcal{U}(\os(6))$.

To that end, we shall introduce a Dirac model where $Pin(6)$ and $\osp(1|2)$ act as a dual reductive pair \cite{Brackx2008} on the eigenfunctions so that their respective irreducible representations can be paired through connections between the Casimir operators.

The Dirac operator $\dir$ as well as $\xir$ and $\euler$ are defined in six dimensions as follows:
\begin{align}\label{eq_dirop}
 \dir=\sum_{\mu=1}^{6}\gamma_\mu\partial_\mu~,\qquad \xir=\sum_{\mu=1}^{6}\gamma_\mu x_\mu~,\qquad \euler=\sum_{\mu=1}^{6}x_\mu\partial_\mu,
\end{align}
with $\{\gamma_\mu,\gamma_\nu\}=-2\delta_{\mu\nu}$. 

These operators have $\osp(1|2)$ as their dynamical algebra. Indeed, with the $\mathbb{Z}_2$-grading involution $\inv$ given by
\begin{align}\label{eq_invop}
 \inv=i^{\nicefrac{6\,}{\,2}}\prod_{\mu=1}^{6}\gamma_\mu,
\end{align}
and
\begin{align}
 J_-=-i\dir,\qquad J_+=-i\xir,\qquad J_0=\euler+3,
\end{align}
the presentation \eqref{eq_osp12} of the $\osp(1|2)$ algebra is realized, with the total Casimir given by
\begin{align}
 \Gamma_{[6]}=\frac{1}{2}\left([J_-,J_+]-1\right)\inv.
\end{align}
It should be noted that to any subset $A\subset[6]$ there corresponds a realization of $\osp(1|2)$. More precisely, if one defines
\begin{align}\label{eq_ospA}
 J_-^{A}=-i\sum_{\mu\in A}\gamma_\mu\partial_\mu~,\qquad J_+^{A}=-i\sum_{\mu\in A}\gamma_\mu x_\mu~,\qquad J_0^{A}=\tfrac{|A|}{2}+\sum_{\mu\in A}x_\mu\partial_\mu,
\end{align}
these generators obey the $\osp(1|2)$ relations \eqref{eq_osp12} with the involution given by
\begin{align}\label{eq_invA}
 \inv^{A}=i^{\nicefrac{|A|\,\,}{\,\,2}}\prod_{\mu\in A}\gamma_\mu
\end{align}
when $|A|$ is even. The Casimir has then the expression
\begin{align}\label{eq_casA}
 \Gamma_{A}=\frac{1}{2}\left([J_-^{A},J_+^{A}]-1\right)\inv^{A}.
\end{align}
We will first couple the six representations pairwise (each pair will correspond to an $\osp(1|2)$ and we can hence use the previous observations for $|A|$ even). We will refer to those pairs using the shortened index notation, $\olin{j}=\olin{1},\olin{2},\olin{3}$. The $\olin{j}$'s will now label the representations we want to pair next, in this two-step process.

The Casimirs associated to one or two such indices are easily calculated using the expression \eqref{eq_casA} and they are found to be
\begin{align}\label{eq_Gammaolinij}
\begin{aligned}
 \hspace{-1em}\Gamma_{\olin{j}}&=-iL_{\olin{j}\,}+\frac{1}{2}\Sigma_{\olin{j}\,},\\
 \hspace{-1em}\Gamma_{\olin{i}\,\olin{j}}&=\big(L_{2i-1,2i}\,\gamma_{2i-1}\,\gamma_{2i}+L_{2i-1,2j-1}\,\gamma_{2i-1}\,\gamma_{2j-1}+L_{2i-1,2j}\,\gamma_{2i-1}\,\gamma_{2j}\\
 &\qquad+L_{2i,2j-1}\,\gamma_{2i}\,\gamma_{2j-1}+L_{2i,2j}\,\gamma_{2i}\,\gamma_{2j}+L_{2j-1,2j}\,\gamma_{2j-1}\,\gamma_{2j}+\tfrac{3}{2}\big)\Sigma_{2i-1,2i}\,\Sigma_{2j-1,2j}.
\end{aligned}
\end{align}
They are immediately recognized as the generators \eqref{eq_Gamma1234} of the commutant of the set $\{J_{12},J_{34},J_{56}\}$ in the spinorial representations of $\mathcal{U}(\os(6))$ that were identified in the previous section.

It is here interesting to point out how the coproduct structure of $\osp(1|2)$ occurs in the Dirac operator.
In two dimensions, the gamma matrices are given in terms of the Pauli matrices:
\begin{align}\label{eq_gamma2d}
 \gamma_1=i\sigma_1,\qquad\gamma_2=i\sigma_2,\qquad \{\gamma_\mu,\gamma_\nu\}=-2\delta_{\mu\nu}.
\end{align}
The involution is simply $\inv=i\gamma_1\gamma_2=\sigma_3$ and coincides with the spin operator which we denoted $\Sigma_{12}$.

We observed that the Dirac equation in $4$D provides an $\osp(1|2)$ made out ot two subsystems each realizing also an $\osp(1|2)$. It must hence result from the coproduct mapping:
\begin{align}
 \dir_{[2]}\mapsto\Delta(\dir_{[2]})=\dir_{[2]}\otimes\inv+1\otimes\dir_{[2]}=\dir_{[2]}^{(1)}\inv^{(2)}+\dir_{[2]}^{(2)}.
\end{align}
This connects with the construction of higher dimensional gamma matrices. Indeed, starting with a realization of the Clifford algebra $\cliff_2$ (generated for example by the $2$ $\gamma_i$'s in \eqref{eq_gamma2d}), a systematic way to construct a realization of a Clifford algebra in two additional dimensions (involving $4$ $\hat{\gamma}_i$'s) is \cite{Gallier2014} to take
\begin{align}
\begin{aligned}
 \hat{\gamma}_1&=\gamma_1\otimes(i\gamma_1\gamma_2)&=&(i\sigma_1)\otimes\sigma_3\\
 \hat{\gamma}_2&=\gamma_2\otimes(i\gamma_1\gamma_2)&=&(i\sigma_2)\otimes\sigma_3\\
 \hat{\gamma}_3&=1\otimes\gamma_1&=&~~\,1\,~\otimes(i\sigma_1)\\
 \hat{\gamma}_4&=1\otimes\gamma_2&=&~~\,1\,~\otimes(i\sigma_2).
\end{aligned}
\end{align}
This construction can be iterated as many times as needed for higher dimensional Clifford algebra realizations in even dimensions. 

With this choice of gamma matrices, the Dirac operator in $4$D reads:
\begin{align}
\begin{aligned}
 \dir_{[4]}&=\gamma_1\partial_1+\gamma_2\partial_2+\gamma_3\partial_3+\gamma_4\partial_4\\
           &=\big(\partial_1(i\sigma_1)+\partial_2(i\sigma_2)\big)\otimes\sigma_3+1\otimes\big(\partial_1(i\sigma_1)+\partial_2(i\sigma_2)\big)\\
           &=\dir_{[2]}\otimes\inv+1\otimes\dir_{[2]},
\end{aligned}
\end{align}
which checks with the expected coproduct result. The algebra involution $\inv$ is realized by the $\sigma_3$ matrix and its occurence is made manifest in this fashion.

\section{Dimensional reduction}\label{sec_dimred}
It is instructive to perform the dimensional reduction of the six-dimensional Dirac operator to $\mathbb{R}^3$. Introduce the cylindrical coordinates
\begin{align}\label{eq_cylindrical}
\begin{aligned}
 x_{2j-1}=\rho_{j}\cos\theta_j,\\
 x_{2j}=\rho_{j}\sin\theta_j,
\end{aligned}\qquad
 j=1,2,3.
\end{align}
We then have transformed expressions for $\dir$, $\xir$ and $\euler$. In particular,
\begin{align}
 \dir=\sum_{j=1}^{3}\left(\obar{\gamma}_{2j-1}\frac{\partial}{\partial\rho_j}+\obar{\gamma}_{2j}\frac{1}{\rho_j}\frac{\partial}{\partial\theta_j}\right),
\end{align}
where
\begin{align}\label{eq_obargamma}
\begin{aligned}
 \obar{\gamma}_{2j-1}&=\phantom{-}\cos\theta_j\,\gamma_{2j-1}+\sin\theta_j\,\gamma_{2j},\\
 \obar{\gamma}_{2j}&=-\sin\theta_j\,\gamma_{2j-1}+\cos\theta_j\,\gamma_{2j}.
\end{aligned}
\end{align}
We can now bring the $\obar{\gamma}_\mu$'s back to their original form (the $\gamma_\mu$'s) by means of a rotation in spin space. Let
\begin{align}\label{eq_Srot}
 S=\prod_{j=1}^{3}\exp\left(-\frac{i\theta_j}{2}\Sigma_{\olin{j}}\right),\qquad\Sigma_{\olin{j}}=i\gamma_{2j-1}\gamma_{2j},
\end{align}
a straightforward calculation shows that
\begin{align}
 S^{-1}\obar{\gamma}_\mu S=\gamma_{\mu},\qquad 1\leq\mu\leq 6.
\end{align}
This rotation however leads to additional terms in the expression of $\dir$, which we can also eliminate with a gauge transformation depending on the radii and of the form \mbox{$e^{B}=\prod_{i=1}^{n}f_i(\rho_i)$}. Requiring that after this additional transformation
\begin{align}\label{eq_dirred}
 \widetilde{\dir}=\sum_{j=1}^{3}\left(\gamma_{2j-1}\frac{\partial}{\partial\rho_j}+\gamma_{2j}\frac{1}{\rho_j}\frac{\partial}{\partial\theta_j}\right)
\end{align}
imposes that
\begin{align}\label{eq_eB}
 e^{B}=\prod_{j=1}^{3}\frac{1}{\sqrt{\rho_j}}.
\end{align}
The following transformation
\begin{align}
 \bo\mapsto\widetilde{\bo}=e^{-B}S^{-1}\bo e^{B}S,
\end{align}
with $e^{B}$ and $S$ given by \eqref{eq_eB} and \eqref{eq_Srot} respectively, is thus to be carried.

The angular momentum $J_{12}$ \eqref{eq_Jij} is one of the elements in the set whose commutant we looked for. It is transformed into
\begin{align}
 J_{12}\mapsto\widetilde{J}_{12}&= e^{-B}S^{-1}\left(-i\frac{\partial}{\partial \theta_1}+\frac{1}{2}\Sigma_{12} \right)S e^{B}\\
            &=-i\frac{\partial}{\partial \theta_1}+\frac{1}{2}\Sigma_{12}+(-i)\left(\frac{-i}{2}\Sigma_{12}\right)=-i\frac{\partial}{\partial \theta_1}
\end{align}
and similar results hold for $J_{34}$ and $J_{56}$.

We also have
\begin{align}
 \xir&\mapsto\widetilde{\xir}=\sum_{j=1}^{3}\rho_j\,\gamma_{2j-1},\label{eq_xirred}\\
 \euler&\mapsto\widetilde{\euler}=\sum_{j=1}^{3}\rho_j\frac{\partial}{\partial\rho_j},\label{eq_eulerred}\\
 \Sigma_{\olin{j}}&\mapsto\widetilde{\Sigma}_{\olin{j}}=\Sigma_{\olin{j}}.
\end{align}
Fixing $\widetilde{J}_{2j-1,2j}\sim k_j$ as a result of separation of variables, we can rewrite
\begin{align}
 \widetilde{\dir}=\sum_{j=1}^{3}\left(\gamma_{2j-1}\frac{\partial}{\partial\rho_j}+\gamma_{2j}\frac{ik_j}{\rho_j}\right).\label{eq_dirredfix}
\end{align}
Note that these reduced operators still generate the same dynamical algebra since this is not altered by conjugation or separation of variables. 

It is now interesting to investigate what is the effect of the reduction on the Casimirs operators. Recall that the reduced Casimir associated to the subset $\{\olin{i_1},\dots,\olin{i_n}\}=A\subset[6]$, with $\olin{j}\equiv\{2j-1,2j\}$, is given by
\begin{align}\label{eq_casAtilde}
 \widetilde{\Gamma}_A=\frac{1}{2}\left([\widetilde{\xir}_A,~\widetilde{\dir}_A]-1 \right)\widetilde{\Sigma}_A,\qquad \widetilde{\Sigma}_A=\prod_{k=1}^{n}\Sigma_{\olin{i_k}}.
\end{align}
The reduced Casimirs:
\begin{centeqn}
\begin{gathered}
 \widetilde{\Gamma}_{\olin{i}},\qquad \widetilde{\Gamma}_{\olin{i}\,\olin{j}},\qquad \widetilde{\Gamma}_{\olin{i}\,\olin{j}\,\olin{k}},
\end{gathered}
\end{centeqn}
will satisfy the Bannai-Ito relations
\begin{align}
 \{\widetilde{\Gamma}_{\olin{i}\,\olin{j}},\,\widetilde{\Gamma}_{\olin{j}\,\olin{k}}\}=\widetilde{\Gamma}_{\olin{i}\,\olin{k}}+2\widetilde{\Gamma}_{\olin{j}}\widetilde{\Gamma}_{\olin{i}\,\olin{j}\,\olin{k}}+2\widetilde{\Gamma}_{\olin{i}}\widetilde{\Gamma}_{\olin{k}}
\end{align}
with
\begin{align}\label{eq_Gammajtilde}
\begin{aligned}
 \widetilde{\Gamma}_{\olin{j}}=\frac{1}{2}\left([\widetilde{\xir}_{\olin{j}},~\widetilde{\dir}_{\olin{j}}]-1\right)\Sigma_{\olin{j}}
 =\widetilde{J}_{\olin{j}}
 =k_j.
\end{aligned}
\end{align}
This should be compared with the system studied in \cite{DeBie2016a}, where the Dirac operator was given in terms of Dunkl derivatives $D_i$ of the $\mathbb{Z}_2$ type
\begin{align}\label{eq_reflect}
 D_i=\partial_i+\frac{k_i}{x_i}(1-R_i),\quad\qquad R_i f(x_i)=f(-x_i).
\end{align}
The Bannai-Ito algebra was also seen to be the symmetry algebra in that case, with the one-index Casimirs $\Gamma_j$ equal to the deformation parameters $k_j$. Here, we started with ordinary partial derivatives in twice as many dimensions and found the one-index Casimirs taking the values of the angular momenta that are diagonalized in the dimensional reduction. Hence, the reduced system obtained here offers a new model with Bannai-Ito symmetry in addition to the Dirac-Dunkl one.

\section{The higher rank Bannai-Ito algebra as a commutant}\label{sec_higher}
Let us now show how one can extend the result of the previous sections to the higher rank Bannai-Ito algebra $B(n)$.

Take any triple of pairwise disjoint subsets of $[2n]$ called $K$, $L$, and $M$. There is an obvious isomorphism
\begin{align}\label{eq_isoosp}
\osp^{K}(1|2)\otimes\osp^{L}(1|2)\otimes\osp^{M}(1|2)\cong\osp(1|2)\otimes\osp(1|2)\otimes\osp(1|2),
\end{align}
so that the Casimir elements $\Gamma_{K}$, $\Gamma_{L}$, $\Gamma_{M}$, $\Gamma_{K \cup L}$, $\Gamma_{K \cup M}$, $\Gamma_{L \cup M}$, and $\Gamma_{K \cup L \cup M}$ will generate $B(3)$ and hence obey
\begin{align}\label{eq_BIKLM}
\{\Gamma_{K \cup L}, \Gamma_{L \cup M}\} = \Gamma_{K \cup M} + 2 \Gamma_{L} \Gamma_{K \cup L \cup M} + 2 \Gamma_{K} \Gamma_{M}.
\end{align}
Now we wish to know $\{\Gamma_{A}, \Gamma_{B}\}$ for any two subsets $A$ and $B$. To that end, take $K = A\setminus B$, $L = A \cap B$, $M = B\setminus A$ to see that in view of \eqref{eq_BIKLM} the corresponding Casimirs satisfy
\begin{align}\label{eq_B(n)}
\{\Gamma_{A}, \Gamma_{B}\} = \Gamma_{(A\cup B)\setminus (A\cap B)} + 2\Gamma_{A\cap B}\Gamma_{A\cup B} + 2 \Gamma_{A\setminus (A\cap B)} \Gamma_{B\setminus (A\cap B)},
\end{align}
which are the structure relations of $B(n)$ given in \eqref{eq_BIn} \cite{DeBie2017}.

Underneath this quick derivation of \eqref{eq_B(n)} is the fact that the entire algebra $B(n)$ is generated by the Casimirs associated to $2$-subsets since the general relations are inferred from those of $B(3)$. Let us map as we have done the $2$-subsets $\{2i-1,2i\}$, $i=1,\dots,n$ of $[2n]$ to the elements $\olin{i}$ of $[\olin{n}]$ so that obviously \mbox{$\{\olin{i},\olin{j}\}\in[\olin{n}]$} corresponds to $\{2i-1,2i,2j-1,2j\}\in[2n]$. It follows that the structure relations for the $4$-subsets in $[2n]$ are those of $2$-subsets in $[\olin{n}]$. Working with the set of $n$ integers viewed in this way and in light of the preceeding remark, it will suffice to examine the generators $\Gamma_{\olin{i}\,\olin{j}}\,$, namely the $\Gamma_{2i-1,2i,2j-1,2j}\,$ of the Dirac model in $2n$ dimensions.

We can now use Howe's duality to conclude that $B(n)$ is the commutant of the subalgebra $\os(2)^{\oplus n}$. Let $\cliff_{2n}$ be the Clifford algebra generated by $\gamma_\mu$, $\mu=1,\dots,2n$ with relations as in \eqref{eq_clifford}. The spatial rotation generators are the $L_{\mu\nu}$ verifying the $\os(2n)$ commutation relations \eqref{eq_Lijrel}. 

Replacing $6$ by $2n$ in equations \eqref{eq_dirop}, \eqref{eq_invop} provides operators $\dir$, $\xir$, $\euler$ and $\inv$ realizing $\osp(1|2)$ in $2n$ dimensions.

Setting $J_-=-i\dir$, $J_+=-i\xir$, $J_0=\euler+n$, the presentation of $\osp(1|2)$ coincides with \eqref{eq_osp12} and the total Casimir is $\Gamma_{[2n]}=\tfrac{1\,}{\,2}\left([J_-,J_+]-1\right)\inv$. The Casimirs associated to one or two labels $\olin{j}$ are easily calculated with the help of formula \eqref{eq_casA} which remains valid for $A\subset[2n]$. One finds exactly the same operators $\Gamma_{\olin{j}}$ and $\Gamma_{\olin{i}\,\olin{j}}$ with expressions as in \eqref{eq_Gammaolinij} for $i,j=1,\dots,n$.

These Casimir operators are immediately recognized as the generators of the commutant of the set $\{J_{12},\dots,J_{2n-1,2n}\}$ in the spinorial representation of $\mathcal{U}(\os(2n))$.

Since we know that the intermediate Casimirs realize the commutation relations of the Bannai-Ito algebra, it is necessarily also the case for the generators of the commutant. This therefore confirms our claim to the effect that $B(n)$ is the commutant of $\os(2)^{\oplus n}$.

We can again perform the dimensional reduction of the $2n$-dimensional model to obtain a $n$-dimensional system whose symmetry algebra is $B(n)$. Write $\dir$ in the cylindrical coordinates \eqref{eq_cylindrical} with $j=1,\dots,n$ and the $\obar{\gamma}_{\mu}$'s defined as in \eqref{eq_obargamma}. Exactly as was done in six dimensions, rotate the $\obar{\gamma}_\mu$'s to their original form with the help of an operator $S$ extending \eqref{eq_Srot} and accompany this with the gauge transformation defined by  $e^{B}=\prod_{j=1}^{n}\rho_j^{-\nicefrac{1\,}{\,2}}$, thus performing \mbox{$\bo\mapsto\widetilde{\bo}=e^{-B}S^{-1}\bo e^{B}S$}. One then obtains for $\widetilde{\dir}$, $\widetilde{\xir}$ and $\widetilde{\euler}$ the same expressions as in \eqref{eq_dirred}, \eqref{eq_xirred} and \eqref{eq_eulerred} with the sum extending to $n$ instead of stopping at $3$.

The angular momenta $J_{2j-1,2j}$ \eqref{eq_Jij} are then mapped to \mbox{$\widetilde{J}_{2j-1,2j}=-i\frac{\partial}{\partial \theta_j}$}.
Fixing $\widetilde{J}_{2j-1,2j}\sim k_j$ once the ignorable variable is eliminated, we can rewrite $\widetilde{\dir}$ as in \eqref{eq_dirredfix} again extending the sum to $n$. 

Note that the reduced operators still generate the same dynamical algebra. 

The reduction of the Casimirs is as described from \eqref{eq_casAtilde} to \eqref{eq_Gammajtilde}, except that $A$ is now a subset of $[2n]$.

The reduced model thus obtained offers a new $n$-dimensional system in addition to the Dirac-Dunkl one, with the $B(n)$ Bannai-Ito algebra as its symmetry algebra. The one-index Casimirs in the different models respectively take the values of the angular momenta and the deformation parameters in the Dunkl-derivatives.

\section{An embedding of the $R(n)$ Racah algebra in the $B(n)$ Bannai-Ito algebra}\label{sec_embedding}
The higher rank Racah algebra $R(n)$ is an associative algebra with generators \mbox{$\{C^{i},P^{ij}\}_{1\leq i<j\leq n}$} and defining relations \cite{DeBie2017a}:
\begin{subequations}\label{eq_Rr}
\begin{align}
 [P^{ij},P^{jk}] &= 2 F^{ijk},\label{eq_Rn1}\\
 [P^{jk},F^{ijk}] &= P^{ik}P^{jk}-P^{jk}P^{ij}+2P^{ik}C^j-2P^{ij}C^k,\label{eq_Rn2}\\
 [P^{kl},F^{ijk}] &= P^{ik}P^{jl}-P^{il}P^{jk},\label{eq_Rn3}\\
 [F^{ijk},F^{jkl}] &= F^{jkl}P^{ij}-F^{ikl}\big(P^{jk}+2C^j\big)-F^{ijk}P^{jl},\label{eq_Rn4}\\
 [F^{ijk},F^{klm}] &= F^{ilm}P^{jk}-P^{ik}F^{jlm},\label{eq_Rn5}
\end{align} 
\end{subequations}
where $i,j,k,l,m\in[n]$ are all different.

In \cite{Gaboriaud2018a} we identified the generators of the commutant of $\os(2)^{\oplus n}$ (in oscillator representations of $\mathcal{U}(\os(2n))$) which are the invariants  $\{G^{i},K^{ij}\}_{1\leq i<j\leq n}$ given in \eqref{eq_GiKij}. With the following redefinitions
\begin{align}\label{eq_Pij}
\begin{aligned}
 C^{i}&=-\frac{1}{4}G^{i}-\frac{1}{4},\\
 C^{ij}&=-\frac{1}{4}K^{ij},\\
 P^{ij}&=C^{ij}-C^{i}-C^{j}=-\frac{1}{4}K^{ij}+\frac{1}{4}\left(G^{i}+G^{j}\right)+\frac{1}{2},
\end{aligned}
\end{align}
a long but straightforward calculation in the oscillator realization showed that the defining relations \eref{eq_Rr} of the algebra $R(n)$ were obeyed.

It has been seen in \cite{Genest2015} that $R(3)$ admits an embedding in $B(3)$. We already noted in Section \ref{sec_B3comm} that the commutant picture brought this inclusion to the fore and we shall exploit it here together with the results in \cite{Gaboriaud2018a} to explicitly provide the embedding of $R(n)$ in $B(n)$.

The key point is that the intermediate Casimirs $C^{i}$ and $C^{ij}$ generating the $R(n)$ algebra can be obtained from the intermediate sCasimirs of the $B(n)$ algebra.

We will only need to use the sCasimirs associated to $2$ or $4$ indices. They are given as follows:
\begin{align}
 S_{\mu\nu}&=\left(L_{\mu\nu}\gamma_\mu \gamma_\nu+\frac{1}{2}\right)\\
 S_{\mu\nu\rho\sigma}&=\left(L_{\mu\nu}\gamma_\mu \gamma_\nu+L_{\mu\rho}\gamma_\mu \gamma_\rho+L_{\mu\sigma}\gamma_\mu \gamma_\sigma+L_{\nu\rho}\gamma_\nu \gamma_\rho+L_{\nu\sigma}\gamma_\nu \gamma_\sigma+L_{\rho\sigma}\gamma_\rho \gamma_\sigma+\frac{3}{2} \right)
\end{align}
and a straightforward calculation in the realization \eqref{eq_Lij} allows to recover the $C^{i}$'s and $C^{ij}$'s as defined in \eqref{eq_Pij} through the following formulae:
\begin{align}
 C^{i}&=\frac{1}{4}\left(S_{2i-1,2i}^{2}-S_{2i-1,2i}-\frac{3}{4}\right),\\
 C^{ij}&=\frac{1}{4}\left(S_{2i-1,2i,2j-1,2j}^{2}-S_{2i-1,2i,2j-1,2j}-\frac{3}{4}\right).
\end{align}
This readily gives the embedding of $R(n)$ inside $B(n)$.

\section{Conclusion}
This paper has offered a novel presentation of the Bannai-Ito algebra $B(n)$ as the commutant of $\os(2)\oplus\dots\oplus\os(2)$ in the spinorial representation of $\os(2n)$ associated to the Clifford algebra $\cliff_{2n}$. It has also indicated how this picture can be elegantly related to the definition of $B(n)$ as the centralizer in $\mathcal{U}(\osp(1|2)^{\otimes n})$ of the coproduct embedding of the Lie superalgebra $\osp(1|2)$ in $\osp(1|2)^{\otimes n}$ in the framework of the Howe duality associated to $\big(Pin(2n)\,,\osp(1|2)\big)$. This called for the introduction of a model involving the (massless) Dirac equation in $2n$ dimensions and identifying the connection between the Casimir operators that describe the paired representations of the two mutually commuting algebras on the solution space.

Invariance under the subalgebra of $Pin(2n)$ allowed for dimensional reduction of the Dirac equation under separation of variables. This resulted in a model in $\mathbb{R}^{n}$ with Bannai-Ito symmetry without reflection operators that hence differ from the Dirac-Dunkl equation already known to possess the same symmetry.

The commutant picture for $B(n)$ made manifest the fact that the Racah algebra $R(n)$ can be embedded in $B(n)$. This observation had been made for $n=3$ \cite{Genest2015} and could here be explicitly extended.

Looking ahead it would be interesting to understand how various contractions of $B(n)$ (and $R(n)$) play out within the commutant presentation. The relation with superintegrable systems would certainly be worth exploring \cite{Kalnins2015,Ruiz2016}.

The Racah algebra is associated to $\sl(2)$ and the Bannai-Ito algebra to $\osp(1|2)$. The Askey-Wilson algebra \cite{Zhedanov1991} is similarly related to the quantum algebra $\mathcal{U}_q(\sl(2))$. The rank $1$ algebra encodes the bispectrality of the Askey-Wilson polynomials. Efforts are now deployed to construct the extensions to arbitrary ranks \cite{Post2017,DeBie2018}. It is natural to think that the Askey-Wilson algebra also admits a dual commutant presentation. We plan on examining this matter which could shed useful light on the higher rank construction. We hope to report on these questions we have raised in the near future.

\ack
The authors wish to thank Jean-Michel Lemay for useful discussions.
JG holds an Alexander-Graham-Bell scholarship from the Natural Science and Engineering Research Council (NSERC) of Canada. 
LV gratefully acknowledges his support from NSERC. 
SV enjoys a Neubauer No Barriers scholarship at the University of Chicago and benefitted from a Metcalf internship.
The work of AZ is supported by the National Foundation of China (Grant No. 11771015).

\section*{References}
\bibliographystyle{unsrtinurl} 
\bibliography{citationsBI.bib}

\appendix
\section{Connection with superconformal quantum mechanics}\label{sec_appendix}
As an aside to our discussion, we wish to observe that the superconformal quantum Hamiltonian with $\sl(2|1)$ symmetry presented in \cite{Fubini1984} can be obtained by dimensional reduction from the two-dimensional harmonic oscillator. 
The Dirac operator in $2$D and the corresponding position and Euler operators are given by
\begin{align}
 \dir=\gamma_1\partial_1+\gamma_2\partial_2,\qquad\xir=\gamma_1 x_1+\gamma_2 x_2,\qquad\euler=x_1\partial_1+x_2\partial_2.
\end{align}
They generate the $\osp(1|2)$ dynamical algebra \eqref{eq_osp12}, precisely realized upon defining \mbox{$J_-=-i\dir$}, \mbox{$J_+=-i\xir$}, \mbox{$J_0=\euler+1$}, and taking the algebra involution to be \mbox{$\inv=\Sigma_{12}=i\gamma_1\,\gamma_2$}. 
The $2$D harmonic oscillator Hamiltonian is the algebra element
\begin{align}
 H_{\text{h.osc.}}=\dir^{2}-\xir^{2}=-\left(\frac{\partial^{2}}{\partial x_1^{2}}+\frac{\partial^{2}}{\partial x_2^{2}}\right)+\left(x_1^{2}+x_2^{2}\right)
\end{align}
and hence possesses this $\osp(1|2)$ symmetry.

Performing the dimensional reduction $(x_1,x_2)\to\rho$, carrying the transformation $e^{B}S$ and fixing the angular momentum to be $J_{12}\sim k$, as explained in section \ref{sec_dimred}, one obtains
\begin{align}
 \widetilde{\dir}=\gamma_1\frac{\partial}{\partial\rho}+\gamma_2\frac{i\,k}{\rho},\qquad\widetilde{\xir}=\gamma_1\rho,\qquad\widetilde{\euler}=\rho\frac{\partial}{\partial\rho},\qquad\widetilde{\Sigma}_{12}=\Sigma_{12}.
\end{align}
With the gamma matrices realized in terms of the Pauli matrices as
\begin{align}
 \gamma_1=i\sigma_1,\qquad\gamma_2=i\sigma_2,\qquad\Sigma_{12}=\sigma_3,
\end{align}
the $2$D harmonic oscillator Hamiltonian is ``reduced'' to
\begin{align}\label{eq_hamfubini}
 \widetilde{H}_{\text{h.osc.}}=-\frac{\partial^{2}}{\partial\rho^{2}}+\frac{k(k-\sigma_3)}{\rho^{2}}+\rho^{2},
\end{align}
which is identified with the superconformal quantum mechanical model introduced and analyzed by Fubini and Rabinovici \cite{Fubini1984}. The supercharges are given by $\widetilde{\dir}$ and $\widetilde{\xir}$. In \cite{Fubini1984} the Hamiltonian \eqref{eq_hamfubini} is actually observed to have the larger $\sl(2|1)$ or $\osp(2|2)$ symmetry. This follows from the fact that $\Sigma_{12}$ is an additional even symmetry which generates two supplementary supercharges when commuted with $\dir$ and $\xir$.

\end{document}